\documentclass[useAMS, usenatbib]{mn2e}

\usepackage[dvips]{graphicx} 
\input{epsf}

\title{Testing the starburst/AGN connection with SWIRE X-ray/70$\mu$m sources}

\author[M. Trichas et al.]
       {M. Trichas$^1$\thanks{Email: m.trichas@imperial.ac.uk},
A. Georgakakis$^2$, 
M. Rowan-Robinson$^1$,
K. Nandra$^1$,
D. Clements$^1$,\newauthor
M. Vaccari$^3$
  \\
        $^1$Astrophysics Group, Blackett Laboratory, Imperial College London,
        Prince Consort Road, London SW7 2BW, UK.\\
        $^{2}$National Observatory of Athens, I. Metaxa $\&$ V. Pavlou, Athens 15236, Greece\\
        $^{3}$Department of Astronomy, University of Padova,
Vicolo Osservatorio 5, I-35122 Padua, Italy}
          \date{Accepted 23 June 2009 }

\pagerange{\pageref{firstpage}--\pageref{lastpage}}
\pubyear{2009}

\begin{document}
\maketitle
\label{firstpage}
\newcommand{\mic}{\umu m}
\newcommand{\gsim}{\mathrel{\rlap{\lower4pt\hbox{\hskip1pt$\sim$}}
    \raise1pt\hbox{$>$}}}   

\begin{abstract}
We  explore  the nature  of  X-ray  sources with  $\rm  70  \, \mu  m$
counterparts selected  in the SWIRE fields ELAIS-N1,  Lockman Hole and
Chandra  Deep   Field  South,  for   which  Chandra  X-ray   data  are
available.  A total  of 28  X-ray/$\rm  70 \,  \mu m$  sources in  the
redshift  interval  $0.5   <  z  <  1.3$  are   selected.   The  X-ray
luminosities  and the  shape  of  the X-ray  spectra  show that  these
sources are  AGN.  Modelling of  the optical to  far-infrared Spectral
Energy Distribution indicates that most  of them (27/28) have a strong
starburst  component  ($ \rm  >  50  \,  M_{\odot} \,  yr^{-1}$)  that
dominates in  the infrared.  It is found  that the X-ray  and infrared
luminosities of the sample  sources are broadly correlated, consistent
with a link between AGN activity and star-formation.  Contrary to the
predictions of some  models for the co-evolution of  AGN and galaxies,
the  X-ray/$\rm 70  \,  \mu m$  sources  in the  sample  are not  more
obscured  at   X-ray  wavelengths   compared  to  the   overall  X-ray
population.  It is also found that the X-ray/$\rm 70 \, \mu m$ sources
have lower specific star-formation  rates compared to the general $\rm
70 \, \mu  m$ population, consistent with AGN  feedback moderating the
star-formation in the host galaxies.

  \end{abstract}

\newcommand{\mnras}{MNRAS}
\newcommand{\apj}{ApJ}
\newcommand{\apjl}{ApJL}
\newcommand{\apjs}{ApJS}
\newcommand{\aj}{AJ}
\newcommand{\aap}{AAP}
\newcommand{\araa}{ARA\&A}
\newcommand{\pasp}{PASP}
\newcommand{\nat}{Nature}

\def\lsim{\mathrel{\rlap{\lower4pt\hbox{\hskip1pt$\sim$}}
    \raise1pt\hbox{$<$}}}     

\section{Introduction}
A major recent development in extragalactic
astronomy is the discovery that a large fraction of the spheroids in
the local Universe contain a super-massive black hole (Magorrian et al. 1998). Moreover, the black hole mass is  tightly
correlated  to the  stellar mass  of  the host galaxy bulge
(eg Ferrarese et al. 2000; Gebhardt et al.  2000), suggesting that
the formation and evolution of galaxies  and the build-up of the
super-massive black holes at their centers are interconnected. The physical
process behind this fundamental observation is under debate, with
AGN-driven feedback mechanisms proposed as a plausible option.  \\
\begin{table*}
\centering
\caption{X-ray/70$\mu m$ sample}
\begin{minipage}{17cm}
\begin{tabular}{lllllllllllllll}
object no. & RA & dec & z\footnote{Photometric redshifts are denoted in brackets}& $L_{cirr}$ \footnote{cirrus luminosity}& $L_{sb}$\footnote{starburst luminosity} & $L_{tor}$\footnote{dust torus luminosity} & $L_{opt}$\footnote{Bolometric optical luminosity} & type\footnote{Optical SED best fit} & $A_V$\footnote{extinction}& $lg L_X$ \footnote{Hard band ($\rm 2-10~keV$) X-ray Luminosity in erg s$^{-1}$} & log N(H) \footnote{Intrinsic column density in cm$^{-2}$}\\
&&&&&&&&&&&&&&\\
{\bf N1}&&&&&&&&&&&\\
1 & 242.03664 & 54.94373 & (0.862) & & 12.70 (A220) & 11.45 & 11.20 & Sbc & & 43.66 & 23.56\\
2 & 242.14854 & 54.39154 & 1.300 & & & 12.73 & 12.82 & QSO & 0.15 &  44.41 & 22.46\\
3 & 242.30495 & 53.90828 & 0.9923 & & 11.87 & 12.17 & 12.32 & QSO & & 45.06 & 21.61\\
4 & 242.45093 & 54.60308 & 0.8730 & & 12.01 & 10.41 & 11.31 & Sbc & 0.2 & 43.84 & 23.67\\
5 & 242.55586 & 55.11386 & (0.542) & & 11.85 (A220) & 11.00 & 11.10 & Sbc & & 42.87 & 22.89 \\
6 & 242.74052 & 54.68533 & (1.270) & & 12.45 & 11.85 & 11.45 & Sbc & & 43.81 & 21.71\\
7 & 242.80289 & 55.13984 & 1.280 & & 12.45 & 11.95 & 11.75 & Sbc & & 44.68 & 23.60\\
8 & 243.26450 & 54.72134 & (1.138) & & 12.34 & 11.20 & 11.61 & Scd & & 43.58 & 20.11\\
{\bf Lockman}&&&&&&&&&&&\\
1 & 161.13074 & 59.09380 & 1.312 & &  12.91 & 12.51 & 12.61 & QSO & 0.10 & 44.09 & 22.79 \\
2 & 161.19244 & 58.82482 & (1.280) & & 12.29  & & 11.35 & Scd & 0.3 & 43.21 & 19.83\\
3 & 161.01555 & 58.98413 & (1.208) & & 12.14 & & 11.14 & sb & 0.6 & 43.18 & 22.38\\
4 & 160.84180 & 58.82229 & (1.051) & & 12.01 & 10.44 & 11.59 & Scd & & 43.43 & 19.83 \\
5 & 160.97850 & 58.98404 & (1.004) & & 12.00 & & 11.25 & Scd & 0.1 & 43.07 & 23.65\\
6 & 161.05058 & 58.91622 & (0.845) & & 11.94 (A220) & & 11.64 & E & & 43.43 & 22.79\\
7 & 161.96295 & 58.86107 & (0.762) & & 12.30 & & 11.60 & E & & 42.81 & 21.87\\
8 & 160.84120 & 59.21511 & (0.754) & & 11.79 & 10.50 & 11.50 & Scd & &41.98 & 21.42\\
9 & 160.85416 & 59.28613 & (0.690) & & 11.58 & 11.08 & 10.98 & Scd & & 43.35 & 23.43\\
10 & 161.76088 & 58.75598 & (0.644) & & 11.70 (A220) & & 10.78 & Scd & 1.0 & 43.35 & 22.38\\
11 & 160.72220 & 59.17268 & (0.614) & 11.96 & & & 11.63 & Scd & 1.2 & 42.94 & 22.37\\
12 & 161.21414 & 59.32958 & 0.463 & & 11.33 & 10.93 & 11.23 & sb & 0.2 & 43.60 & 22.45\\
13 & 161.03317 & 58.74367 & 0.555 & & 11.76 & 11.56 & 11.10 & Sbc & 0.2 & 42.25 & 22.54\\
14 & 161.24121 & 58.56060 & (0.905) & & 12.11 (A220) & 10.44 & 11.46 & sb & & 42.99 & 22.56\\
15 & 161.27065 & 58.99651 & 0.264 & & 10.57 & & 9.77 & Sab & & 42.83 & 23.74\\
{\bf CDFS}&&&&&&&&&&&\\
1 & 52.84087 & -27.85638 & (1.032) & & 12.42 (A220) & 12.32 & 11.02 & sb & & 43.20 & 23.30\\
2 & 52.87523 & -27.93405 & (0.660) & & 11.98 (A220) & & 11.53 & Scd & 0.2 & 41.74 & 19.85\\
3 & 52.95019 & -27.80056 & (1.080) & & 12.33 (A220) & 11.13 & 10.33 & sb & & 42.25&19.84\\
4 & 53.01069 & -28.05587 & (1.158) & & 12.24 & 11.64 & 11.06 & Scd & &  43.60 & 19.89\\
5 & 53.02176 & -28.07097 & (0.675) & & 11.58 & 11.32 & 10.90 & Scd & & 42.98 & 22.48\\
\end{tabular}
\end{minipage}
\end{table*}
Recent simulations which include  AGN feedback and
postulate that AGN form during galaxy mergers (Hopkins et al. 2005; 2007) have received  much
attention  recently  because  of   their  success 
in reproducing   a range   of   observed   properties   of   both
AGN  (e.g. duty-cycle,  $\rm N_H$ distribution; Hopkins et al. 2005) and  galaxies (e.g.
bimodality of  the colour-magnitude relation; Blanton 2006). In this scenario galaxy
interactions  trigger the AGN and also produce nuclear starbursts that
both feed and obscure the  central engine for most of its active
lifetime. AGN-driven outflows (i.e. feedback) eventually develop,
which at  later 
stages become  strong   enough  to  rapidly   quench  the
star-formation, leaving behind a red, passive remnant and allowing the
AGN to shine unobscured for a short period ($\approx 10^{8}$\,yr).\\ 
A prediction of  these simulations is  that there should  be an
association  between  AGN  and  starburst events.   Examples  of  such
composite objects are rather  isolated in the literature however (Cid Fernandes 
et al.  2001; Page et al.  2004; Alexander et al. 2005),
particularly  at  $z  \approx 1$,  close  to  the  peak of  the  AGN
luminosity   density in the Universe (Barger  et   al.   2005;
Hasinger  et al. 2005). This observational fact may point  to a more  complex
AGN/starburst connection than that adopted in current models. For
example,  observations show that  X-ray    and/or  optically-selected
AGN   are often found  in red early-type bulge-dominated galaxies (Grogin et al. 2002;  
Pierce et al. 2006; Nandra et al.  2007) with
evidence for star-formation events that have terminated in the recent
past  (Kauffmann et  al. 2003 but see Silverman et al. 2009). These observations are consistent
with a time  lag between the peak  of the star-formation and the BH
accretion in these systems. \\ 
Alternatively, observational  effects may bias our understanding
of  the properties of  AGN hosts.   A major  problem in  assessing the
level of  star-formation in AGN  is the difficulty in  decomposing the
stellar  from  the  AGN  emission,  especially in  the  case  of  dust
enshrouded   systems.   Combining  X-ray   data  with   longer  far-IR
wavelengths (e.g.  $\rm 70\mu m$)  can provide a handle on this issue.
The X-rays provide  the most efficient way of  identifying active BHs,
least  biased by  obscuration, while  common  wisdom has  it that  the
far-IR  is a clean  star-formation diagnostic. The sample, which is by no
 means complete, presented in this paper, has been compiled based on 
the above criteria: objects close to the peak of the AGN luminosity density 
which have both X-ray and 70$\mu m$ detections, represents the best 
candidates for composite objects. Because of instrumental limitations 
(e.g. sensitivity, confusion), any sample selected at $\rm > 70\mu m$, 
like the one proposed here, represents the brightest and  rarest  examples  
of an  underlying population that future more sensitive surveys with Herschel will unveil.
\section{Data}
The sample of X-ray sources with $\rm 70\mu m$ counterparts studied in
this paper is compiled from  the SWIRE survey (Lonsdale et al.  2003),
one of the  largest Spitzer legacy programmes. It  covers a total area
of $\rm 49 deg^2$ spread in  6 different fields.  The observations have
been carried out in all  seven photometric bands available to Spitzer,
the IRAC 3.6, 4.5, 5.8 and 8$\rm \, \mu m $ bands (Fazio et al.  2004)
and the MIPS 24, 70 and 160$\,  \rm \mu m$ bands (Rieke et al.  2004).
The typical 5$\sigma$ sensitivities is  about 3.7, 5.3, 48, 37.7$\mu Jy$
in the IRAC 3.6, 4.5, 5.8 and 8$\rm \, \mu m $ bands respectively. The
corresponding 5$\sigma$ limits  for the MIPS bands are  230$\mu Jy$ at
24$\mu$m and 20, 120$mJy$ at 70  and 160$\rm \, \mu m$ (Surace et al.,
2005).  Optical observations in  at least 3 photometric bands  are available 
for most  of the SWIRE area ($>$70$\%$).
The optical  data combined with  the SWIRE mid- and  far-IR photometry
have  been  used  to fit  the  observed  SEDs  with a  combination  of
galaxy and AGN templates (Babbedge et al.  2004; Rowan-Robinson et al.
2005) in  order to study  the nature of infrared  selected populations
and  to estimate  photometric  redshifts for  over  1 million  sources
(Rowan-Robinson et  al.  2008).  Details  about the latest  SWIRE data
release can be  found in Surace et al.  (2005). In  this paper we have
used multiwavelength data from three SWIRE fields that have X-ray
data available from Chandra.  These are the ELAIS-N1, the Chandra Deep
Field South (CDFS) and the Lockman field.  The overlap  between the Chandra
X-ray observations and the SWIRE survey in these three fields is about
$\rm 2.5~deg^{2}$. Details about individual fields are listed below.  
\begin{figure}
\begin{center}
\includegraphics[width=7.5 cm]{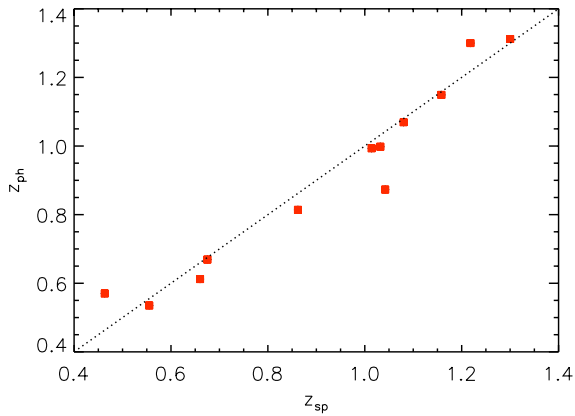}
\caption{Photometric against  spectroscopic reshift estimates  for the
X-ray/$\rm 70\mu m$ sources. Photometric redshifts are estimated using
the SED fitting method  described in the text. Spectroscopic redshifts
are from either our own follow-up program (Trichas et al. in prep) or from the literature. The
diagonal line corresponds $z_{phot}=z_{spec}$.}\label{fig_zz}
\end{center}
\end{figure}
\subsection{ELAIS-N1}
The entire  $\rm 9 \, deg^{2}$  area of SWIRE ELAIS-N1  field has been
observed at optical  wavebands as part of the  Wide Field Survey (WFS,
McMahon et al.,  2001) using the Wide Field Camera  (WFC) on the Isaac
Newton  Telescope (INT). The  survey consists  of 600\,s  exposures in
five bands:  U, g', r', i' and  Z to magnitude (Vega,  5$\sigma$ for a
point-like  object)  limits  of:   23.4,  24.9,  24,  23.2,  21.9\,mag
respectively. The overall photometric accuracy of the INT
WFS survey is 2$\%$. Further details are given in Babbedge et al. (2004)
and Surace et al. (2005).\\
\begin{figure}
\begin{center}
\includegraphics[width=8.5 cm]{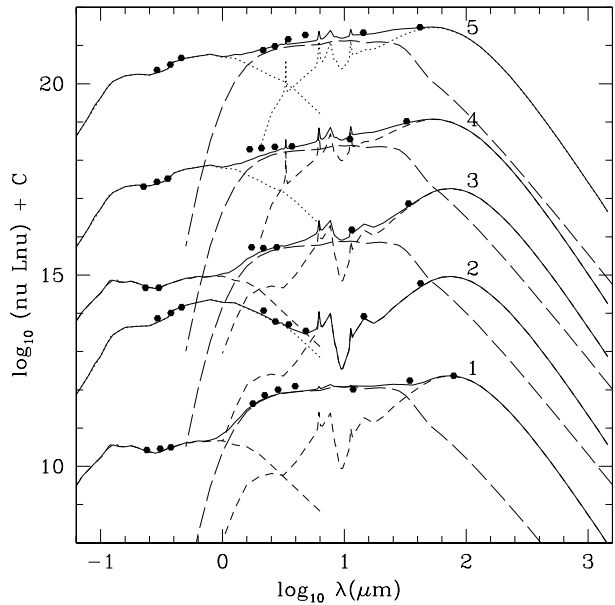}
\caption{Optical to far-infrared SEDs in $\nu f_\nu$ of all 5 X-ray sources with 
70$\mu$m from CDFS. Solid curves show the total predicted SED. Parameters 
for model fits given in Table 1.}
\end{center}
\end{figure}
X-ray coverage of ELAIS-N1 consists of two Chandra surveys.  The first
one is  our own,  and consists of  30$\times$5\,ks exposures  with the
ACIS-I instrument.   There is significant  overlap between
adjacent ACIS-I pointings, thereby providing nearly even coverage over
a circular area of about $\rm 1.5\, deg^{2}$. This survey has not been
fully  presented in  a previous  publication, except briefly in
Georgakakis et al.  (2008), and  therefore we will describe it in some
detail here.   The data were obtained  in VFAINT mode  during Cycle 6,
between  December  2005  and  March  2006.   The  data  were  analysed
following the methodology of Laird  et al.  (2009). The CIAO v3.2 data
analysis software  was used for the data  reduction.  After correcting
for   hot  pixels   and   cosmic  ray   afterglows   using  the   {\sc
acis\_run\_hotpix}  task, the level  2 event  file was  produced using
{\sc acis\_process\_events}. Flares  were identified and removed using
the  automated method described  by Laird  et al.   (2009). Individual
event files although largely  overlapping are treated separately to
avoid  problems  arising   from  merging  regions  with  significantly
different PSFs.  Images are constructed in four energy bands 0.5--7.0,
0.5--2.0, 2.0--7.0  and 4.0--7.0\,keV.  Energy  weighted exposure maps
for each  of the  four energy bands were constructed  using the
{\sc merge\_all}  task by adopting a power-law  spectrum with spectral
index $\Gamma=1.4$ to estimate the energy weights.  For each of the 30
Chandra pointings of the survey, source catalogues were constructed in
the  energy bands above  adopting a  Poisson probability  threshold of
$<4\times10^{-6}$,  which are then  merged into  a master  source list
following the  methodology of  Laird et al.   (2009).  In the  case of
duplicate sources in the overlap regions of adjacent pointings we keep
the detection with the smallest off-axis angle. The count rates in the
0.5--7.0, 0.5--2.0, 2.0--7.0 and  4.0--7.0\,keV bands are converted to
fluxes  in the standard  0.5--10, 0.5--2,  2--10 and  5--10\,keV bands
respectively   assuming  a   power-law  X-ray   spectrum   with  index
$\Gamma=1.4$ and Galactic absorption $\rm N_H=2\times10^{20}\,cm^{-2}$
appropriate for the ELAIS-N1 field.  The limiting fluxes are estimated
$1.0\times10^{-14}$, $3.4 \times 10^{-15}$ and $1.5\times 10^{-14} \,
\rm  erg^{-1}\,s^{-1} \,cm^{-2}$  in the  0.5-10, 0.5-2  and 2-10\,keV
bands respectively. A total of 545  unique sources  are detected
over $\rm  1.5\, deg^{2}$ in  at least one of  the four
energy bands to the flux limits above.
\noindent The second Chandra survey of the ELAIS-N1 is that presented by Manners
et  al.  (2003; 2004).   This consists  of  a single  75\,ks Chandra  ACIS-I
pointing,  which  lies in  middle  of  our  circular geometry  shallow
survey.   Here we  use the  X-ray source  catalogue produced  by these
authors.   They   detect  127  unique   sources  to  flux   levels  of
$2.3\times10^{-15}$, $9.4 \times 10^{-16}$ and $5.2\times 10^{-15} \,
\rm erg^{-1}\,s^{-1} \,cm^{-2}$ in the 0.5-8, 0.5-2 and 2-8\,keV bands
respectively.

\subsection{Chandra Deep Field South}
\begin{figure*}
\begin{center}
\includegraphics[width=8.5 cm]{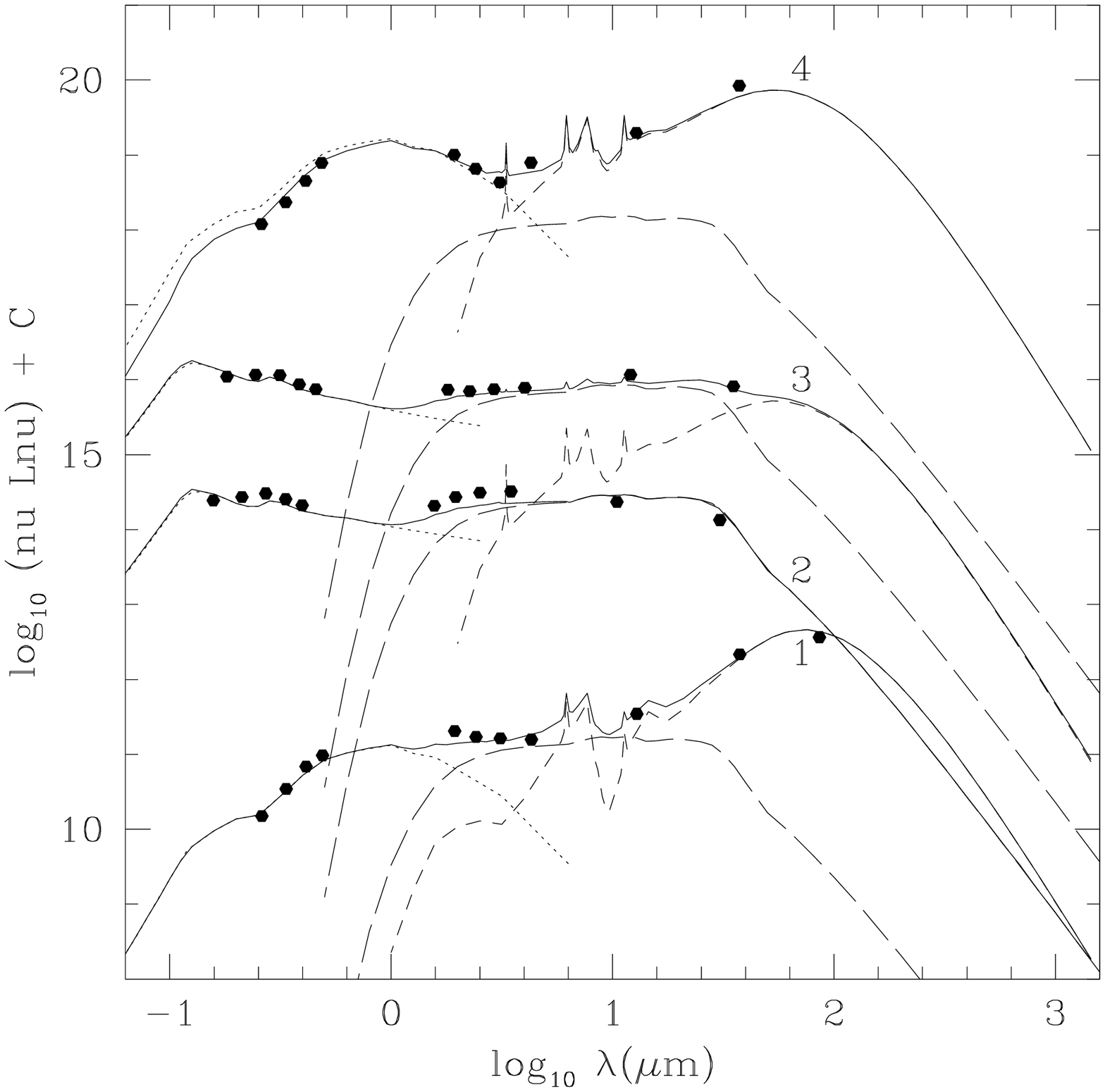}
\includegraphics[width=8.5 cm]{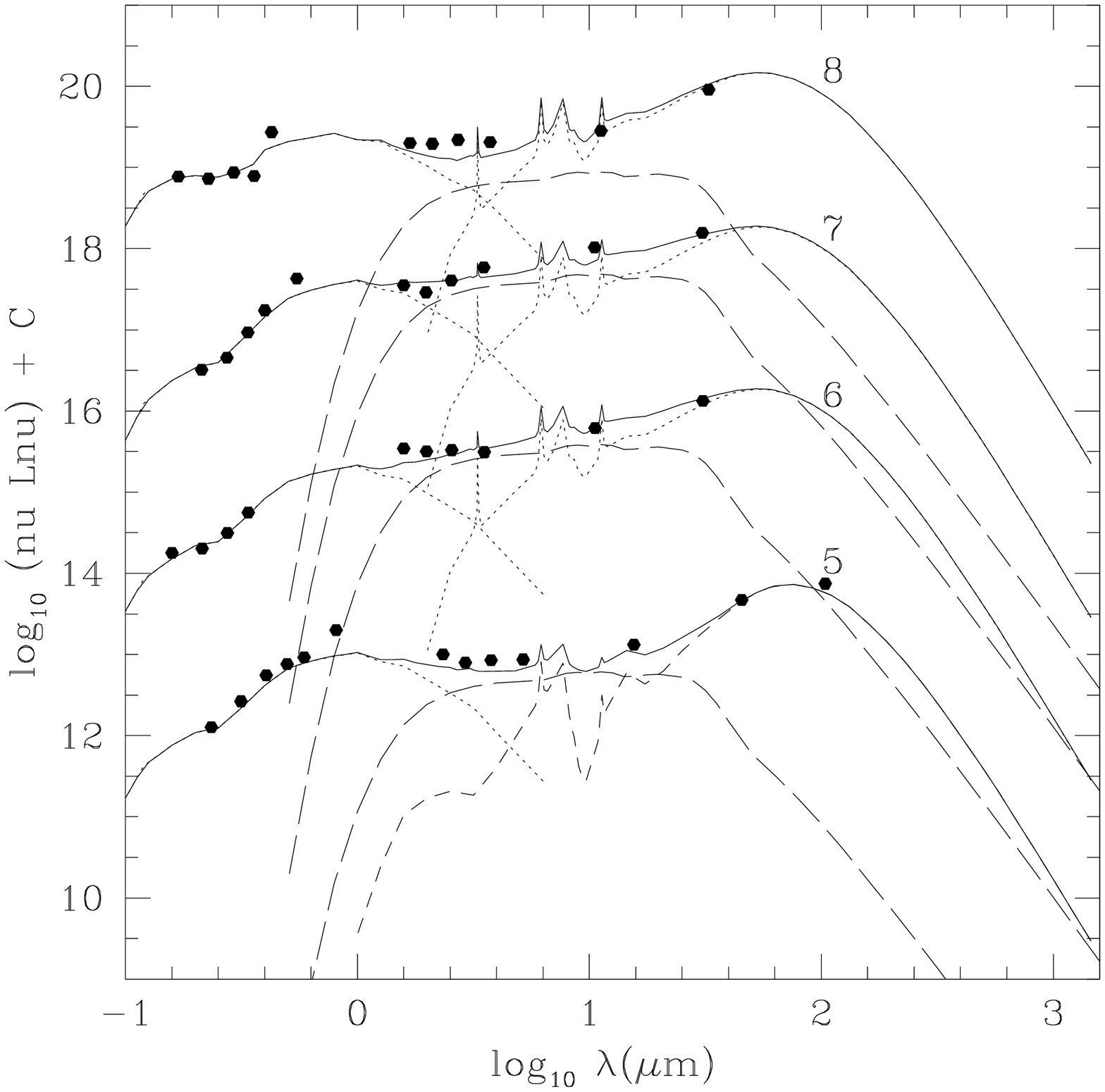}
\caption{Optical to far-infrared SEDs in $\nu f_\nu$ of all 8 X-ray sources with 70$\mu$m from ELAIS-N1. Solid curves show the total predicted SED. Parameters for model fits given in Table 1.}
\end{center}
\end{figure*}
\begin{figure*}
\begin{center}
\includegraphics[width=8.5 cm]{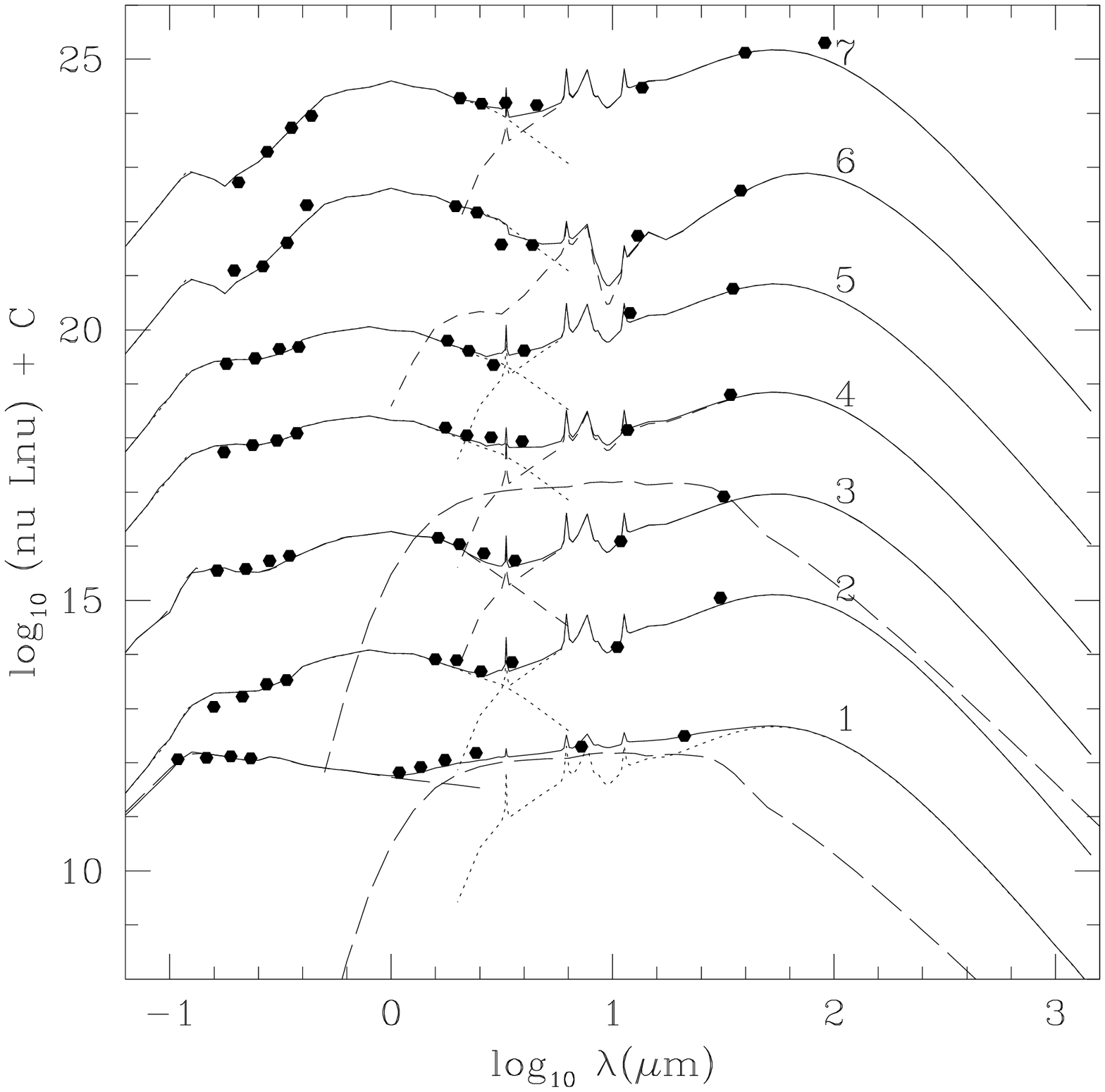}
\includegraphics[width=8.5 cm]{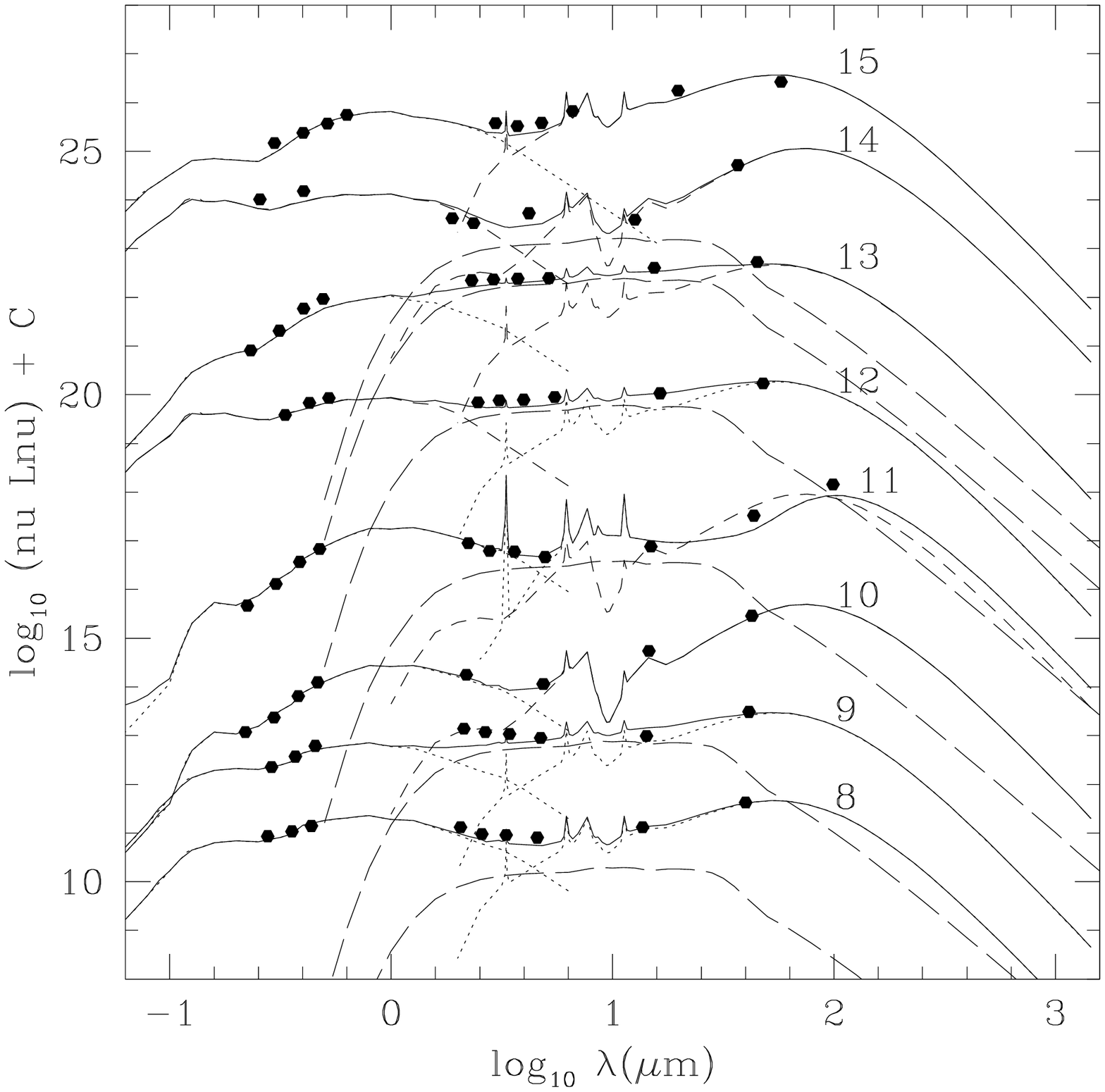}
\caption{Optical to far-infrared SEDs in $\nu f_\nu$ of all 15 X-ray sources with 70$\mu$m from Lockman. Solid curves show the total predicted SED. Parameters for model fits given in Table 1.}
\end{center}
\end{figure*}
The  SWIRE CDFS  was observed  with  the MOSAIC  II camera  on the  4m
telescope at Cerro  Tololo. Fifteen  pointings covered $\rm \sim
4.5 \,  deg^{2}$ in four  filters, U, g,  r, i to 5$\sigma$  depths of
24.5, 25.4,  25, 24 magnitudes (Vega) respectively  with an additional
$\rm 1.5\,  deg^{2}$ survey in the  Z filter down  to 23.3 magnitudes.
There is also a deeper $\rm 0.33 \, deg^{2}$ pointing in U, g, r, i to
5$\sigma$  depths of  25.2, 25.7,  25.5, 24.5  magnitudes respectively (Lonsdale et al., 2003).\\
The X-ray data are from the  Extended CDFS (ECDFS) and the 1\,Ms CDFS.
We use  the source catalogues of  Lehmer et al.  (2005)  for the ECDFS
and Giacconi  et al. (2002) for  the CDFS.  The  Extended Chandra Deep
Field  South  consists  of  4  ACIS-I pointings  covering  $\rm  0.3\,
deg^{2}$ with an exposure time of  $\rm 250 \,ks$ each. A total of 592
sources  are   detected  by  Lehmer  et  al.    (2005)  which  reaches
sensitivity limits of  $1.1 \times 10^{-16}$, $\rm 6.7\times10^{-16} \,
erg  \,  s^{-1}  \,  cm^{-2}$   in  the  0.5--2  and  2--8\,keV  bands
respectively.  The  ECDFS is flanking the 1\,Ms  CDFS survey. Giacconi
et al. (2002)  have detected 316 X-ray sources down  to flux limits of
$5.5\times10^{-17}$,  $\rm  4.5\times10^{-16}  \,  erg  \,  s^{-1}  \,
cm^{-2}$ in the 0.5--2 and 2--10 keV bands respectively.

\subsection{Lockman Hole}

The Lockman Hole field was observed in the U, g', r', and
i' bands with the MOSAIC Camera at the Kitt Peak National
Observatory (KPNO) Mayall 4m Telescope, February 2002 (g',
r', and i') and January 2004 (U band). The scale of the Camera
is 0.26''/pix and the field of view is 36' x 36'. The astrometric
mapping of the optical MOSAIC data is good to less than 0.4"
and the seeing varied between 0.9 and 1.4 arcsec. Data reduction was performed with the Cambridge Astronomical Survey
Unit (CASU, Irwin $\&$ Lewis 2001) pipeline, following the procedures described in Babbedge et al (2004). Fluxes were measured
within a 3" aperture (diameter) and corrected to total fluxes using
growth curves. Typical 5 sigma magnitude limits are 24.1, 25.1,
24.4 and 23.7 in U, g', r' and i' respectively (Vega), for point-like sources (Berta et al 2007a).\\
The X-ray data are from the  medium deep Chandra survey of Polletta et
al  (2006). It  consists of  9 ACIS-I  contiguous pointings  of 70\,ks
each  which  cover an  area  of  $\rm  0.6\,deg^2$. The  catalogue  of 
Polletta  et al  (2006)  has a  total  of 827  unique  sources to  the
0.3--8\,keV  flux  limit  of  about  $10^{-15}  \, \rm  erg  \,  s  \,
cm^{-2}~$. \\
The total number of Chandra X-ray sources in the three fields is 2091.
 
\section{The X-ray/70 $\mu$m sample}
\begin{figure}
\begin{center}
\includegraphics[width=8.5 cm]{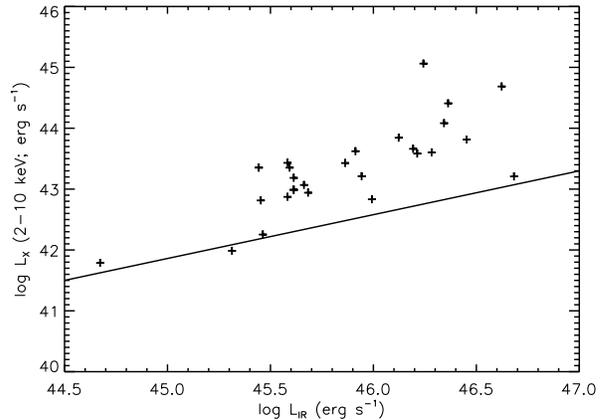}
\caption{2-10\,keV  X-ray  luminosity  as   a  function  of  total  IR
luminosity,   $L_{IR}$.  Black crosses  are  the X-ray/$\rm  70\,   \mu  m$
sources. The  continuous line is the X-ray/IR  luminosity relation for
local  star-forming galaxies from  Ranalli et  al. (2003).  
All X-ray/$\rm 70\, \mu m$ sources deviate  from this Ranalli et  al. relation in  the 
sense that they  are  X-ray luminous  for  their  $L_{IR}$,  indicating that  AGN
activity dominates the X-ray luminosity.}\label{fig_lxlir} 
\end{center}
\end{figure}
In order to  identify X-ray sources
with  70\,$\mu  m$ counterparts  we  have  cross-correlated the  SWIRE
catalogue with the X-ray source lists described above following the same procedure discussed in Fadda et al. (2002), Franceschini et al. (2005) and Polletta et al. (2006).  The total number of X-ray/SWIRE sources with 24 $\mu m$ detections is 1368. In order to explore  the possible link between AGN activity and star-formation  at the  epoch close to  the peak of  AGN luminosity
density (e.g. Barger  et al. 2005), we further  restrict the sample  to the redshift  interval $0.5<z<1.3$ using  either  spectroscopic (if  available)  or photometric  redshift
estimates.  This resulted in a total of 28 X-ray/SWIRE sources with 70$\mu m$ detections. Given the surface density of X-ray and 70 $\mu m$ sources within this redshift interval, the expected number of random associations is $<2$.
This sample consists of 5 CDFS
sources, 1 from the  1 Ms survey and 4 from the  ECDFS, 8 are ELAIS-N1
sources, 7  from the 5ks survey and  one from the EDXS  and 15 sources
are in Lockman. Table 1  gives the positions, redshifts 
(bracketed if a photometric redshift) and rest-frame luminosities for the different
optical, infrared and X-ray components fitted to these sources.

\subsection{Multiwavelength Properties}\label{sec_sed}
In order  to classify the  SEDs and estimate photometric  redshifts of
the sources  in our  sample, the $U$-band to  $\rm 4.5\, \mu$m photometric data 
are first fitted using a  library of  14 templates,  3 QSO, 1  Starburst and  10 galaxy
templates  (Rowan-Robinson  et al.  2008).  Photometric redshifts  are
estimated at this step.  At  the longer Spitzer wavebands, the stellar
contribution  is  first  subtracted   from  the  photometric  data  by
extrapolating  the best  fitting  galaxy template  from the  previous
step.  The corrected 3.6-170 $\mu$m data are then fitted with a mixture of four templates,
cirrus (quiescent),  M82 or Arp220 starbursts,  and  AGN dust torus,  providing  information 
on  the nature  of sources  and  the dominant  mechanism  responsible for  the
infrared emission, AGN vs stellar processes. The availability of $\rm
70 \, \mu m$ detections for all 28 sources and $\rm 160 \, \mu m$
detections for 7  of  them  is   an  advantage,  providing  additional
leverage for elucidating the infrared properties of the sample. \\
For the sample of 28 X-ray/$\rm 70\mu m$ sources the minimum number of
optical/near-IR  bands  (up  to $\rm  4.5  \,  \mu  m$) used  for  the
photometric redshift calculation is 5.  This is sufficient to estimate
reliable  photometric  redshifts.   This  is  demonstrated  in  Figure
1, which plots photometric against spectroscopic redshift for
a  total  of  12  sources  in the  sample  with  optical  spectroscopy
available either from  our own follow-up programs (Trichas  et al.  in
prep.)  or from the literature.  The   agreement  is  good,  with  an
estimated  accuracy of  $\delta z/(1+z_{spec})\approx0.02$  for  the
photometric  redshift  determinations   of  the  X-ray/$\rm  70\mu  m$
sources.  Broad-band SEDs for all 28 sources are given in Figures 2, 3, 4.
The optical galaxy or AGN fits to the $U$-band to  $\rm 4.5\, \mu$m data
show that the photometric redshift solutions are plausible.  All but two of
the 28 sources require an M82 or A220 starburst component.  The two exceptions are
N1-2, which appears to have only an AGN dust torus component in the infrared,
and Lockman-11, which seems to be better fitted with a quiescent ('cirrus')
component.  19 of the 28 sources require an AGN dust torus component.
Since we believe the X-ray emission of all but two of the 28 galaxies is due to an AGN (see
section 4), the remaining 7 galaxies must be cases with weak dust tori
(see discussion by Rowan-Robinson et al. (2009) of the Lockman SWIRE-CLASX
sample).  
\section{Results}
\begin{figure}
\begin{center}
\includegraphics[width=7.8 cm]{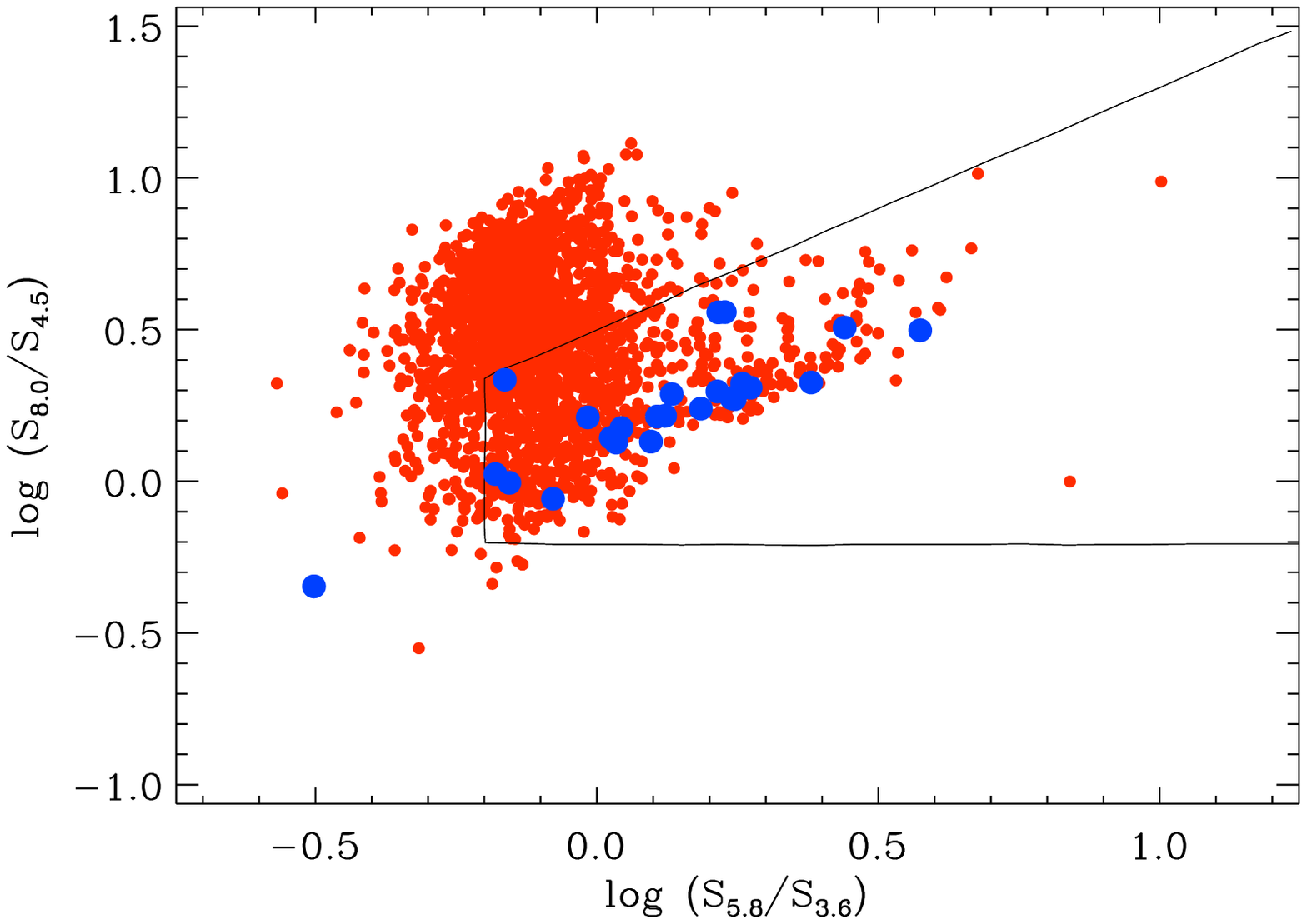}
\caption{IRAC/MIPS colour-colour plot. Solid line represents the predicted 
obscured AGN parameter space (Lacy et al., 2004). Red dots are SWIRE 70$\mu$m sources within the redshift interval 0.5-1.3. Blue dots are the 28 X-ray/70$\mu$m sources.}\label{a_plot}
\end{center}
\end{figure}The  X-ray  and mid-IR properties  of  the   X-ray/$\rm  70  \,  \mu  m$  sources
indicate that most of them host an AGN. Figure 5 shows
that the  X-ray/$\rm 70 \, \mu  m$ sources are offset  to higher X-ray
luminosities compared  to the  Ranalli et al.   (2003) $L_X  - L_{IR}$
correlation  for  starbursts.  Only two sources in the sample, Lockman-7
and CDFS-2, may possibly be X-ray starbursts, since they have $L_X/L_{ir}$ ratios
only slightly higher than the Ranalli et al (2003) mean, and show no evidence for
AGN activity (infrared dust torus, broad emission lines, QSO optical SED). Also,  the  obscuration corrected  X-ray
luminosities  of 26  out of  28 sources  in the  sample are  $L_X (\rm
2-10\,keV) \ga 10^{42} \, erg  \, s^{-1} \, cm^{-2}$, higher than what
is typically attributed to  star-formation (Georgakakis et al.  2007).
The X-ray  spectra of  19 X-ray/$\rm  70 \, \mu  m$ sources  are hard,
consistent with  column densities $N_{H}  \ga 10^{22} \rm  \, cm^{-2}$
(assuming  a power-law  intrinsic X-ray  spectrum  with $\Gamma=1.9$),
thereby indicating  an obscured AGN.  At  infrared  wavelengths the  X-ray/$70\,  \rm  \mu  m$ sources  have
luminosities $L_{IR} > 10^{11} \,  L\odot$ and therefore belong to the
class  of  Luminous   and  Ultra-Luminous  Infrared  Galaxies  (LIRGs,
ULIRGs). The template fitting method of section 3 provides
information on  the emission  mechanism(s) that are responsible  for these
bright  luminosities.   The  mid-IR  part of  their  SEDs  include a significant
contribution (up to  90 per cent at  $\rm 8\, \mu m$) from  a hot dust
component associated with AGN  activity. This  is   further  demonstrated   in  Figure 6 which shows the position  of the X-ray/$\rm 70 \, \mu m$
sources on the mid-IR colour-colour  plot of Lacy et al.  (2004).  The
majority are  in the  region of the  parameter space occupied  by QSOs
(Georgantopoulos et  al. 2007) and  power-law selected AGN  (Donley et
al. 2008).\\
\begin{figure}
\begin{center}
\includegraphics[width=7.8 cm]{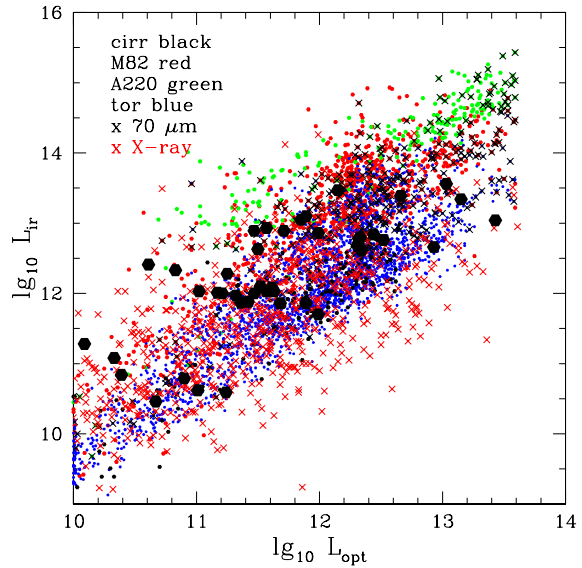}
\caption{Bolometric infrared luminosity, $L_{ir}$, versus 0.1-3.0 $\mu$m optical luminosity.
The small filled circles are SWIRE 24 $\mu$m QSOs (black cirrus, red M82, 
green A220, blue AGN dust torus, data from Rowan-Robinson et al 2008). The black crosses are the subset of 
these with 70 $\mu$m detections. The red crosses are X-ray/SWIRE sources with 24 $\mu$m detections (1368 
objects).  X-ray luminosities are calculated as in RR, Valytchanov and 
Nandra 2009.  The large black filled hexagrams are the subset of these with 70 $\mu$m 
detections.}

\end{center}
\end{figure}However, in
addition to the  contribution from AGN heated dust,  our SED modelling
also  suggests  a  star-formation   component  in  the  mid-IR in almost all cases.   This
component   becomes   increasingly   important  at   longer   infrared
wavelengths and is essential to fit  the the 70 and/or $\rm 160 \, \mu
m$ emission of  27 out  of 28 sources  in the  sample. 
Quantitatively,  according to  our template  fits,  for 26  out of  28
sources in Table 1, $>50$ per cent of the $L_{IR}$ is from cool dust
associated with  star-formation. The
estimated  SFRs are $\ga  50 \,  \rm M_{\odot}/yr$,  thereby indicating
intense starburst  events. In summary,  the multiwavelength properties
of the X-ray/$\rm 70 \, \mu m$ sources indicate they are systems where
both the stellar population and the SBH are growing at the same time. It
is interesting to  explore whether these two processes  are related in
the sample sources. \\
For optically selected QSOs for example, Netzer et
al.  (2007) and Serjeant $\&$ Hatziminaoglou (2009) found  that the $\rm 60\mu m$  luminosity, which estimates
the current star-formation rate, scales with the 5100\,\AA\, continuum
luminosity, which  is a proxy of  the total AGN  power, suggesting the
two processes  are linked. Figure 7 shows an equivalent plot, the
bolometric  infrared luminosity  ($L_{IR}$; $\rm  3-1000\, \mu  m$) to
approximate the luminosity of  the current star-formation rate versus the
optical bolometric  luminosity, $L_{opt}$ ($\rm 0.1-3\, \mu  m$) for all SWIRE QSOs with 24 $\mu$m detections (data from Rowan-Robinson 
et al 2008), as a
proxy  of  the  AGN power (see  figure  caption  for details  on  the
estimation  of luminosities). X-ray/$\rm 70  \, \mu  m$  sources and
SWIRE  QSOs  show the  same  broad  correlation  between $L_{IR}$  and
$L_{opt}$ suggesting an association  between the formation of stars in
these galaxies  and the growth of  the SBH at their  centres.  We note
however,  that  the  correlation  in  Figure 7  has
substantial  scatter, i.e.  at  a given  $L_{opt}$ the  $L_{IR}$ spans
about 2\,dex. We have also shown the corresponding distribution for the whole sample of 
SWIRE/X-ray sources with 24 $\mu$m associations (red crosses in Fig 7), with X-ray luminosities first converted
to a bolometric luminosity assuming a bolometric correction to the hard X-ray luminosity
of 27, and then to $L_{opt}$ by dividing the bolometric luminosity by 2
(Rowan-Robinson et al 2008).  X-ray galaxies with very high values of 
$L_{sb}/L_{opt}$ are likely to be X-ray starbursts.\\
In contrast with the 70 $\mu$m QSOs, optically selected samples like the PG QSO 
samples of Netzer et al (2007) and Lutz et al (2008)
are biased towards lower values of $L_{sb}/L_{opt}$.  Optical selection favours
objects with higher $L_{opt}$, 70 $\mu$m selection favours higher $L_{ir}$. The 70$\mu$m/X-ray sample (filled hexagrams in Fig 7) seem to cover the whole range of
$L_{opt}$ and $L_{ir}$
\\
 \begin{figure}
\begin{center}
\includegraphics[width=7.8 cm]{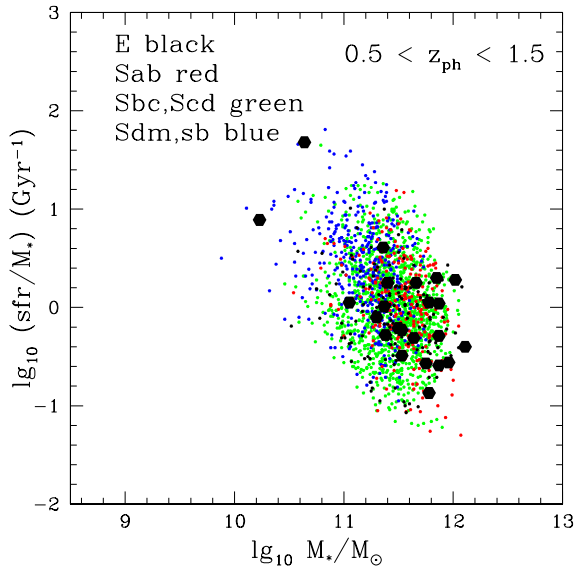}
\caption{Specific star formation rate in $Gyr^{-1}$ versus stellar mass in $M_{\odot}~yr^{-1}$
for SWIRE galaxies, colour-coded by optical template type.  the 70$\mu$m/X-ray sources
are overplotted as large black filled hexagrams. 
Apart from two objects (CDFS-1 and CDFS-3) at the top left, there is a 
tendency for our sample to have somewhat high stellar masses ($>2x10^{11}$$M_{\odot}$) 
and somewhat lower specific star formation rates ($<$3$Gyr^{-1}$ ). Both the two low stellar 
mass objects in our sample have dust tori.}
\end{center}
\end{figure}Figure 8 shows the specific star-formation rate versus stellar mass for the entire SWIRE $\rm 70 \, \mu m$  population, in the
redshift  interval  0.5-1.5,
colour-coded by optical template type,
overplotted by our sample. Apart from two objects, CDFS-1 and CDFS-3 with low
stellar masses and high specific star-formation rates, the 70$\mu$m/X-ray
population is at the lower end of the specific star-formation rate distribution  for 0.5 $<$ z $<$ 1.5, 
and tends to have a  higher stellar  masses
than the general  population, i.e. they avoid the
most intense starbursts of similar stellar mass.\\
It is also  interesting that a large fraction of  the X-ray/$\rm 70 \,
\mu  m$ sample  (19/28) are  obscured by  column densities  $N_{H} \ga
10^{22} \rm  \, cm^{-2}$.  Figure 9 compares the HR  distribution of
the X-ray/$\rm  70 \, \mu  m$ sources with  that of the  overall X-ray
population with SWIRE photometric  redshifts in the interval 0.5--1.5.
A Kolmogorov-Smirnov  tests shows that the likelihood  of the observed
differences  if  the  two  samples  are drawn  from  the  same  parent
population is about 3 per cent. Therefore there is no strong evidence
(3 per cent  or about $2\sigma$ in the case  of a normal distribution)
that the  X-ray/$\rm 70 \, \mu  m$ sources are more  obscured than the
overall X-ray population.
 \section{Discussion}
 \begin{figure}
\begin{center}
\includegraphics[width=9. cm]{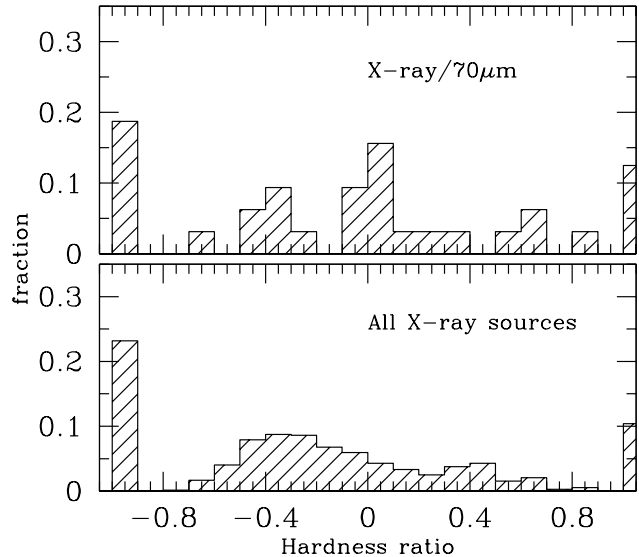}
\caption{Hardness ratio distribution for X-ray/$\rm  70 \, \mu  m$ sources and the entire X-ray 
source population identified with SWIRE sources in the redshift interval. 
The hardness ratio is defined as HR=(H-S)/(H+S) where H is the hard band 
(2-7keV) count rate and S is the soft band (0.5-2 kev) count rate. The KS 
test shows that the probability that the distribution are drwan from the 
same parent samples is 3$\%$, i.e. there are no statisitically significant 
($<$2$\sigma$) differences between the two populations.}
\end{center}
\end{figure}Chandra  X-ray data  within the  SWIRE  survey footprint  are used  to
select a total  of 28 X-ray sources with detections at  $\rm 70 \, \mu
m$,  with additional
160$\mu$m for 25$\%$ of them, in the  redshift interval 0.5-1.3  to study the  properties of
galaxies that are  best candidates for systems where  the formation of
the stellar population is taking place  at the same time as the growth
of the SBH at their centres.  Study of the multiwavelength $U$-band to
far-IR  SED of these  systems indeed  shows that  the majority  of the
sample harbors powerful starbursts  (27/28; $\ga  50 \,  \rm M_{odot}/yr$), while
the X-ray data indicate AGN activity in at least 26/28 sources. \\
The  X-ray  spectra of  19/28  sources  are  consistent with  moderate
obscuration, $\rm N_H \ga  10^{22} \, cm^{-2}$.  However, the obscured
AGN fraction in the sample is similar to that of the overall X-ray AGN
population in the same redshift interval. This contradicts some models
for the co-evolution  of AGN and galaxies in  which the central engine
is  obscured by  dust  and gas  clouds  associated with  circumnuclear
starbursts  (e.g.   Fabian  et   al.   1999;  Hopkins  et  al.   2006;
Ballantyne 2008).  Alternatively,  our finding  may indicate
that a substantial fraction of  the AGN has ongoing star-formation and
that the X-ray/$\rm 70 \, \mu  m$ might not be very different in their
star-formation properties  from the bulk of the  X-ray AGN population.
Silverman et  al.  (2009)  for example, used  the [OII]\,3727  line as
proxy of the star-formation rate  in X-ray selected AGN at $z\approx1$
and showed that the majority show evidence for star-formation.\\
We  also find  that X-ray/$\rm  70 \,  \mu m$  avoid the  most intense
starbursts at  a given  stellar mass. This  is consistent  with models
where outflows from the  central engine moderate the star-formation by
sweeping away the cold gas reservoirs of the host galaxy (e.g. Hopkins
et al.  2006).  This result also contradicts Silverman  et al.  (2009)
who  finds  that X-ray  selected  AGN in  the  COSMOS  field have,  on
average, higher  star-formation rate compared to non  X-ray sources in
that field of similar stellar mass.
\section{Conclusion}
SWIRE infrared and optical data  have been combined with Chandra X-ray
observations  in the  ELAIS-N1, Lockman  Hole and  CDF-S to  study the
nature of the X-ray/$\rm 70 \,  \mu m$ sources in the redshift interval 0.5-1.3.  The majority of these
systems are starburst/AGN composites thereby providing the opportunity
to explore the relation between  SBH growth and galaxy formation.  The
X-ray/$\rm 70 \,  \mu m$ sources follow the  broad correlation between
star-formation rate  (approximated by the $L_{IR}$)  and SBH accretion
rate (approximated by $L_{opt}$)  for broad line QSOs, suggesting that
these  two processes  are linked.  We do  not find  evidence  that the
X-ray/$\rm  70 \,  \mu m$  are more  obscured than  the  overall X-ray
population,  contrary   to  the  predictions   of  some  AGN/starburst
co-evolution  models.  Also, the  X-ray/$\rm 70  \, \mu  m$ population
appears to  avoid the most  intense starbusts for their  stellar mass.
This may indicate  AGN driven moderation of the  star-formation in the
host galaxy.
\section{ACKNOWLEDGMENTS}
The authors wish to thank Elise Laird and James Aird for their assistance in reducing Chandra data and the SWIRE team for their constructive comments on the original Chandra ELAIS-N1 proposal. This work has been supported by funding from the Peren grant (MT) and the STFC Rolling grant (MT).

\label{lastpage}
\end{document}